\documentclass[a4paper,11pt]{article}
\usepackage{pos}
\usepackage[capitalise]{cleveref}
\def\Dz{D^0\to K^-\pi^+}
\def\Dp{D^+\to K^-\pi^+\pi^+}
\title{Precise measurements of $D$ meson lifetimes}

\author*[1]{N.~K.~Nisar }

\affiliation[a]{Brookhaven National Laboratory,\\
  Upton, New York, 11973}

\note{On behalf of the Belle II Collaboration}
\emailAdd{nnellikun@bnl.gov}

\abstract{We report the result of $D^0$ and $D^+$ lifetime measurement using $\Dz$ and $\Dp $ decays reconstructed using $72~{\rm fb^{-1}}$ of data collected by the Belle II experiment at SuperKEKB asymmetric-energy $e^{+}e^{-}$ collider. The results, $\tau(D^0)=410.5\pm1.1({\rm stat})\pm0.8({\rm syst})~{\rm fs}$ and $\tau(D^+)=1030.4\pm4.7({\rm stat})\pm 3.1({\rm syst})~{\rm fs}$, are the most precise to date and are consistent with previous measurements.
}

\FullConference{%
  *** Particles and Nuclei International Conference - PANIC2021 ***\\
  *** 5 - 10 September, 2021 ***\\
  *** Online ***
}

\begin{document}
\maketitle

\section{Introduction}
Accurate predictions of charm meson lifetimes are challenging due to the contributions of strong-interaction to the decay amplitudes and it is an important ingredient to many theoretical calculations as well as experimental measurements. The predictions must resort to effective models, such as the heavy-quark expansion~~\cite{Neubert:1997gu,Uraltsev:2000qw,Lenz:2013aua,Lenz:2014jha,Kirk:2017juj,Cheng:2018rkz} and precise lifetime measurements provide excellent tests of such models. The lifetime measurement with early Belle II data will demonstrate the excellent vertexing capability of the Belle II detector which is essential for future analyses of decay-time-dependent effects.

In this paper, we report the measurement of $D^0$ and $D^+$ lifetimes by reconstructing $D^{*+}\to(\Dz)\pi^+$ and $D^{*+}\to(\Dp)\pi^0$ decays using ${\rm 72~fb^{-1}}$ of data collected by Belle II detector~\cite{belle2}(Charge-conjugate decays are implied throughout). The $D^{*+}$ tagging is requested to suppress the combinatoric background. In SuperKEKB~\cite{skekb}, the $D^{*+}$ mesons are produced with a boost that displace the $D^0$ and $D^+$ mesons. The decay time is estimated from the projection of this displacement on to the direction of momentum, $\vec{p}$, as $t=m_{D}\vec{L}\cdot\vec{p}/|\vec{p}|^2$, where $m_D$ is the known mass of the relevant $D$ meson~\cite{pdg}. The uncertainty in decay time, $\sigma_t$, is estimated by propagating the uncertainties in $\vec{L}$ and $\vec{p}$, including their correlations.      

\section{Belle II detector}

The Belle II detector is built around the interaction region of SuperKEKB $e^+e^-$ collider. 
The inner most part is a two-layer silicon-pixel detector (PXD) followed by a four-layer double-sided silicon-strip detector (SVD) and a central drift chamber (CDC) together form the tracking system. A time-of-propagation counter and an aerogel ring-imaging Cherenkov counter that cover the barrel and forward end-cap regions of the detector, respectively, are used for charged-particle identification. An electromagnetic calorimeter is used to reconstruct photons and electrons. All these components are kept inside a 1.5 T magnetic field. A   dedicated system to identify $K^0_L$ mesons and muons is installed in the outermost part of the detector.     

\section{Reconstruction}

\begin{figure}[t!]
\centering
\includegraphics[width=0.5\linewidth]{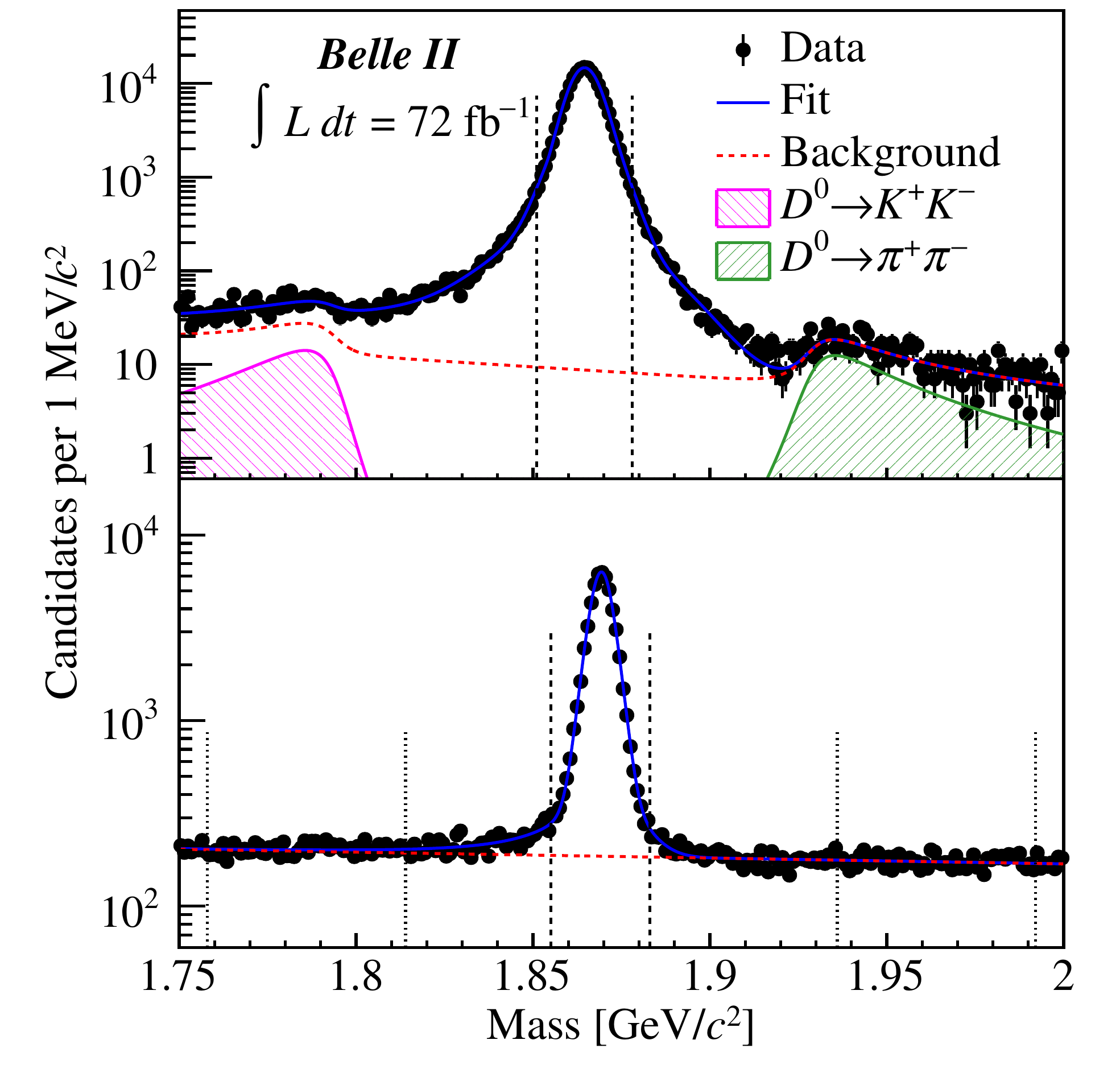}\\
\caption{Mass distributions of (top) $\Dz$ and (bottom) $\Dp$ candidates with fit projections overlaid. The vertical dashed and (for the bottom plot) dotted lines indicate the signal regions and the sideband, respectively.\label{fig:massfit}}
\end{figure}

$\Dz$ and $\Dp$ candidates are reconstructed using charged tracks identified as kaons and pions. Each track is required to have at-least one hit in the first layer of PXD, one hit in the SVD. Tracks from $D^0(D^+)$ need to have at least 20 (30) hits in the CDC. The low momentum $\pi^+$ from the $D^{*+}$ decay are tracks consistent with originating from the interaction region that have at least one hit in the SVD and one hit in the CDC. Low momentum $\pi^0$ is reconstructed from two photons as $\pi^0\to\gamma\gamma$. The $D^{*+}$ momentum in $e^+e^-$ centre-of-mass frame is required to be greater than $2.5(2.6)~{\rm GeV}/c$ to suppress $D^0(D^+)$ mesons coming from bottom mesons. A global decay-chain vertex fit~\cite{vxf}  constraining the tracks according to the decay topology is applied and only candidates with fit $\chi^2$ probabilities larger than 0.01 are retained for further analysis. The mass of $D^0$ and $D^+$ candidate is required to be $1.75<m(K^-\pi^+)<2.00~{\rm GeV}/c^2$. The difference between the $D^{*+}$ and $D$ candidate masses, $\Delta M$, must satisfy $144.94 < \Delta M < 145.90~{\rm MeV}/c^2$ and $138 < \Delta M < 143~{\rm MeV}/c^2$ for $D^0$ and $D^+$ candidates, respectively. By applying these selections, approximately $171\times10^3$ signal $D^0$ candidates with signal purity of 99.8\% is observed in the signal region, defined as $1.815<m(K^-\pi^+)<1.878~{\rm GeV}/c^2$. The signal region in $m(K^-\pi^+\pi^+)$ is defined as $1.855<m(K^-\pi^+\pi^+)<1.883~{\rm GeV}/c^2$ and contains approximately $59\times10^{3}$ signal candidates with a background contamination of 9\%. Mass distributions of $\Dz$ and $\Dp$ candidates are shown in Fig.~\ref{fig:massfit}.

\section{Lifetime extraction}

Lifetimes are extracted by using unbinned maximum-likelihood fits to the $(t,\sigma_t)$ distributions of candidates populating the signal regions. The signal probability-density function (PDF) is the convolution of an exponential function in $t$ with a resolution function that depends on $\sigma_t$, multiplied by the PDF of $\sigma_t$. The time constant of the exponential function will return the lifetime. The PDF of $\sigma_t$ is a histogram template derived directly from the signal region of the data. In the $D^0$ case, the PDF of $\sigma_t$ is obtained assuming that all candidates in the signal region are signal decays. In the $D^+$ case, instead, the template is obtained from the candidates in the signal region after having subtracted the distribution of the sideband data. Simulation shows that a double (single) Gaussian with common mean will describe the resolution function for $D^0(D^+)$. The mean of the resolution function is allowed to float in the fit to account for a possible bias in the determination of the decay time; the width is the per-candidate $\sigma_t$ scaled by a free parameter $s$ to account for a possible misestimation of the decay-time uncertainty. 

In the $D^0$ case, the per-mille-level fraction of background candidates in the signal region is neglected and a systematic uncertainty is assigned for this. A sizable background contamination is accounted for in the $D^+$ case using the data sideband: $1.758<m(K^-\pi^+\pi^+)<1.814, 1.936<m(K^-\pi^+\pi^+)<1.992~{\rm GeV}/c^2$. The background PDF consists of a zero-lifetime component and two exponential components, all convoluted with a Gaussian resolution function having a free mean and a width corresponding to $s\sigma_t$. To better constrain the background
parameters, a simultaneous fit to the candidates in the signal region and sideband is performed by constraining the background fraction obtained from a fit to $m(K^-\pi^+\pi^+)$. 

\begin{figure}[t!]
\centering
\includegraphics[width=0.5\linewidth]{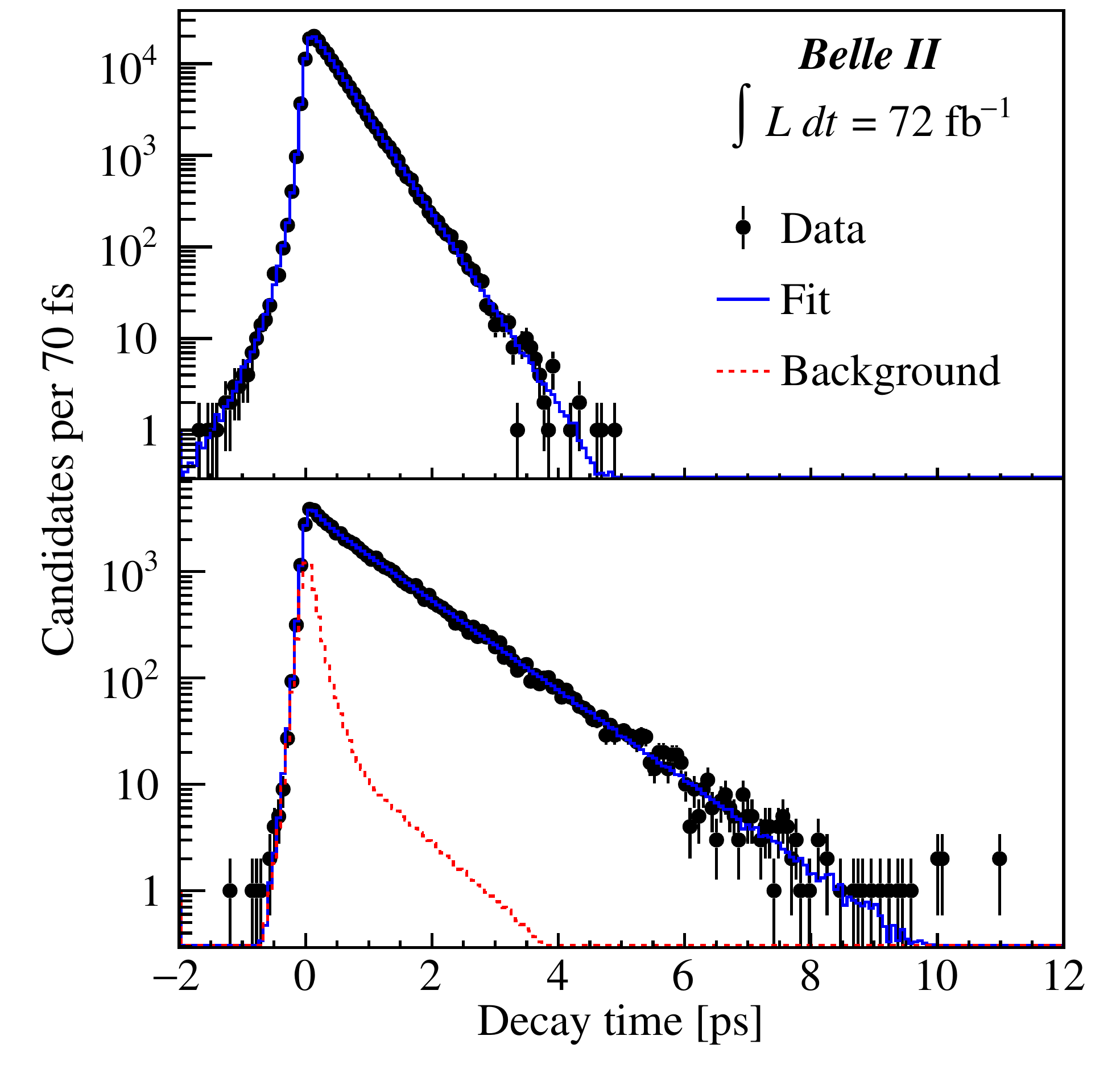}\\
\caption{Decay-time distributions of (top) $\Dz$ and (bottom) $\Dp$ candidates in their respective signal regions with fit projections overlaid.\label{fig:lifetime-fit}}
\end{figure}

The lifetime fits are tested on simulated samples and the returned lifetimes are consistent with the true values.  The decay-time distributions of the data, with fit projections overlaid, are shown in Fig.~\ref{fig:lifetime-fit}. The $D^0$ and $D^+$ lifetimes are measured to be $410.5\pm 1.1(\rm stat)\pm0.8 (\rm syst)$ fs and $1030.4\pm4.7(\rm stat)\pm3.1(\rm syst)$ fs, respectively~\cite{dl_prl}. The results are consistent with their respective world average values~\cite{pdg}. The systematic uncertainties are summarized in Table~\ref{tab:syst} and the total systematic uncertainty is the sum in quadrature of the individual components.

\begin{table}[t]
\centering
\caption{Systematic uncertainties.\label{tab:syst}}
\begin{tabular}{lcc}
\hline
Source & $\tau(\Dz)$ [fs] & $\tau(\Dp)$ [fs]\\
\hline
Resolution model   & 0.16 & 0.39 \\
Backgrounds        & 0.24 & 2.52 \\
Detector alignment & 0.72 & 1.70 \\
Momentum scale     & 0.19 & 0.48 \\
\hline
Total  & 0.80 & 3.10 \\
\hline
\end{tabular}
\end{table}
\section{Systematic Uncertainty}

A small correlation between $t$ and $\sigma_{t}$ is neglected in our nominal fitting model. In order to quantify the effect 1000 signal-only samples of simulated events with same statistics as data are fitted with the nominal PDF. Upper bounds of 0.16 fs and 0.39 fs on the average absolute deviation of measured lifetimes from their true value is assigned as a systematic uncertainty due to imperfect resolution for $\Dz$ and $\Dp$, respectively.  

A background contamination of 0.2\% is neglected in the signal region of $\Dz$. To estimate the effect on our result, 500 simulated samples of $e^+e^-$ events with same size and signal-to-background ratio as data are fitted with the nominal model. The average absolute deviation of fitted lifetime, after subtracting the uncertainty due to resolution modeling, from the true value is 0.24 fs and assigned as systematic uncertainty due to background contamination.  

The background in $\Dp$ signal region is modeled using data sideband. A mismatch between data and simulation in the sideband may be indicating an imperfect description of background components in signal region by the sideband. 1000 samples prepared using pseudo experiments in signal region and simulated data in sideband reproduce the same level of disagreement is fitted and the absolute average difference between the measured and simulated lifetime, 2.52 fs, is assigned as the systematic uncertainty due to background modeling.

Misalignment of tracking detectors may cause bias in the decay-length determination and hence the lifetime. Two sources of uncertainties associated with the alignment are considered: the statistical precision and a possible systematic bias. The day-to-day difference between alignments in real data is used for the statistical contribution. Samples of same statistics as data are simulated by introducing realistic misalignment effects and the difference in lifetime residual for a given misalignment configuration and that from a perfectly aligned sample is assigned as systematic uncertainty.

\section{Conclusions}
In conclusion, the $D^0$ and $D^+$ lifetimes are measured using the data collected by the Belle II experiment corresponding to an integrated luminosity of $72~{\rm fb}^{-1}$. The results are the most precise to date and are consistent with previous measurements.

\end{document}